\documentstyle[12pt,psfig]{article}
\def \m{{\hat F}_{-}}
\def \p{{\hat F}_{+}}
\def \phit{\tilde{\phi}}
\begin{document}

\begin{titlepage}
\thispagestyle{empty}
\begin{flushright}
UMD-GR97-38\\
\end{flushright}
\vskip 1cm
\begin{center}
{\LARGE\bf Computing the spectrum of black hole radiation in the
presence of high frequency dispersion: an analytical approach}\\
\vskip 1cm
{\large Steven Corley}
\vskip .5cm
{\it Department of Physics, University of Maryland\\
                          College Park, MD 20742-4111, USA}\\
             {\tt corley@physics.umd.edu}\footnote{current
email address: scorley@phys.ualberta.ca}

\end{center}
\vskip 1cm
              
\begin{abstract}
We present a method for computing the spectrum of black hole
radiation of a scalar field satisfying a wave equation with high
frequency dispersion. The method involves
a combination of Laplace transform and WKB techniques for finding approximate
solutions to ordinary differential equations.  
The modified wave equation is
obtained by adding
a higher order derivative term suppressed by powers of a fundamental
momentum scale $k_0$ to the ordinary wave equation.  Depending on 
the sign of this new term, high
frequency modes propagate either superluminally or subluminally.  We
show that the 
resulting spectrum of created particles is thermal at the Hawking
temperature, and further that the out-state is a thermal state at
the Hawking temperature,
to leading order in $k_0$, for either modification.

\end{abstract}
\end{titlepage}
\vskip 2mm

\section{Introduction}

Since Hawking's discovery that black holes radiate a thermal spectrum,
\cite{Hawk75}, various other derivations of this effect have appeared,
\cite{HartHawk76,Unruh76,FredenHaag}.  All seem to 
depend in some crucial way on the very
high frequency behavior of the theory.  Clearly such
a derivation cannot be trusted without taking into account backreaction
effects from the spacetime metric.  This has led a number of authors,
\cite{Unruh95,BMPS,Jac,CJ}, to consider the effects of 
high frequency dispersion on the Hawking spectrum.  All have found,
in the context of specific models, that the Hawking radiation remains almost
exactly thermal at the Hawking temperature.

In this paper we show how the leading order contribution to the Hawking
flux can be obtained by analytical methods for two models containing
high frequency dispersion, one in which the high frequency modes
propagate subluminally and one where the high frequency modes propagate
superluminally.  The method involves a combination of
WKB and Laplace transform techniques to solve the modified wave
equations, similar techniques were also used by Brout, Massar, Parentani,
and Spindel \cite{BMPS} for different subluminal type models.  We 
show for both models that the Hawking flux remains
exactly thermal at the Hawking temperature to leading order in inverse
powers of $k_0$.  We further show, to leading order
in $k _0$, that static observers far outside
the black hole see the in-vacuum as a thermal
state at the Hawking temperature, also in agreement with the
ordinary wave equation case \cite{Wald75}. 

The specific models of high frequency dispersion
considered in this paper are obtained by adding a higher derivative term,
suppressed by a new fundamental momentum cutoff $k_0$, to the ordinary
wave equation with the appropriate sign to generate either subluminal
or superluminal propagation of high frequency modes.
The low frequency modes however behave as in the
ordinary wave equation.
Various subluminal theories have already been considered in \cite{Unruh95,
BMPS,Jac,CJ}, in particular the subluminal equation considered in this paper
is the same as considered in \cite{CJ}.
Recently Unruh \cite{Unruhpc} has also considered a superluminal 
modification (although different from the one considered here)
to the ordinary wave equation.  He has shown 
by numerically solving the modified
wave equation that the spectrum is very nearly thermal at the 
Hawking temperature, and that the Hawking particles arise from vacuum
fluctuations inside the horizon.  Since these vacuum fluctuations
would in principle evolve out of the singularity, a boundary condition
at the singularity would be required.  To avoid this problem, Unruh
instead demanded that the vacuum fluctuations were in the ground state
outside the singularity.
In this paper we consider
a similar superluminal modification to the ordinary wave equation, with
the same type of boundary condition.

The remainder of this paper is as follows.  We begin by introducing the model
in section \ref{model}.  In section \ref{computing} we discuss the method
used
to compute the particle creation in this model.  Sections \ref{sub}
and \ref{sup} then describe how the relevant solutions for
particle creation are obtained for the subluminal and superluminal
dispersion relations respectively, and in section \ref{conclusions}
we end with some conclusions.  We use units with $c=\hbar=1$.

\section{Model}
\label{model}

We consider a real scalar field propagating in a 2-dimensional 
black hole spacetime with metric 
\begin{equation}
ds^2 = - dt^2 + (dx - v(x) dt)^2.
\label{metric}
\end{equation}
This is a generalization of the Lema\^{\i}tre line element of
Schwarzschild spacetime where 
$v(x) = - \sqrt{2 M/x}$.  We follow the convention that $v(x) \leq 0$
and (in units where $c = 1$) the horizon is located at $v(x_h) = -1$.  The
action for the field is given by
\begin{equation}
S = \frac{1}{2} \int d^2 x \, \left[ \left( (\partial_t 
+ v \partial_x) \psi \right)^2
+ \psi {\hat F}(\partial_x) \psi \right].
\label{action}
\end{equation}
To motivate this action we note that the black hole defines a preferred
frame, the frame of freely falling observers.  In the Lema\^{\i}tre 
coordinate
system, $(\partial_t + v(x) \partial_x)$ is the unit tangent to free fall
observers who start from rest at infinity, and $\partial_x$ is
its unit, outward pointing normal.  Our action comes from modifying
the derivative operator only along the unit normal $\partial_x$.

\begin{figure}[hbt]
\centerline{
\psfig{figure=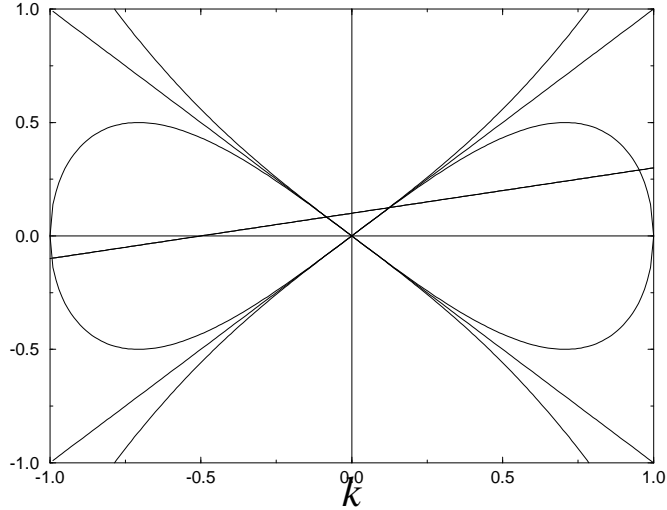,angle=-90,height=8cm}}
\caption{\small Plot of the dispersion relations for the ordinary
wave equation and its subluminal and superluminal modifications.
The intersection points with the $(\omega - v k)$ line are the
possible wavevector roots.  For the particular $v$ shown, there are
four real wavevector roots for the subluminal dispersion relation, but
only two real roots for the superluminal case.  The other two roots
in this case are complex.}
\label{dispgraph}
\end{figure}

In the ordinary, minimally coupled action, ${\hat F}(\partial_x) = 
\partial_{x}^{2}$.  In this paper we take
\begin{equation}
{\hat F}_{\pm}(\partial_x) = \partial^{2}_x \pm \frac{1}{k_{0}^{2}}
\partial^{4}_x.
\label{derivop}
\end{equation}
The essential difference between these two derivative operators is
that ${\hat F}_+$ produces subluminal propagation of high frequency 
modes\footnote{Actually
${\hat F}_+ $ produces both subluminal and superluminal propagation,
however, the superluminal behavior does not play a crucial role in the
analysis of this paper.  It could in fact be removed by considering
a model like Unruh's, \cite{Unruh95}.}  
whereas ${\hat F}_-$ produces superluminal propagation. 
The simplest way to see
this behavior is to look at the respective dispersion relations.
Varying the action produces the equations of motion
\begin{equation}
(\partial_t+\partial_x v)(\partial_t+v\partial_x)\phi=
\partial_{x}^{2} \phi \pm \frac{1}{k_{o}^{2}} \partial_{x}^{4} \phi.
\label{eom}
\end{equation}
Assuming for simplicity that $v(x)$ is constant, we solve 
(\ref{eom}) by looking
at mode solutions of the form
\begin{equation}
\phi(t,x) = \exp(i(\omega t - k x)).
\end{equation}
This produces the dispersion relations
\begin{equation}
(\omega - v k)^2  = k^2 \mp (k/k_0)^4.
\label{dispreln}
\end{equation}
In figure \ref{dispgraph} we plot the square root of (\ref{dispreln}), that
is we plot $(\omega - v k)$ and $\sqrt{k^2 \pm k^4/k_{0}^2}$ as 
functions of $k$ for
a fixed $\omega$,
along with the square root of the dispersion relation resulting 
from the ordinary
wave equation.   
The intersection points are the allowed wavevector roots to (\ref{dispreln}).
The value of the slope of the $\sqrt{k^2 \pm k^4/k_{0}^2}$ 
curve
evaluated at
an intersection point is the locally measured (by a freely falling
observer)
group velocity of a wavepacket
centered about that wavevector.  From figure \ref{dispgraph} we see
for the $\sqrt{k^2 + k^4/k_{0}^2}$ (corresponding to ${\hat F}_-$) curve
that this slope is approximately one when $k \ll k_0$ and increases
with increasing $k$, therefore the high frequency modes propagate
superluminally.  Similarly the $\sqrt{k^2 - k^4/k_{0}^2}$ 
(corresponding to ${\hat F}_+$) curve has
slope approximately one when $k \ll k_0$ and decreases with increasing
$k$ until at some finite $k$ it goes to zero, with a further increase
in $k$ the magnitude of the slope increases and eventually becomes much
larger than one.  Therefore the ``large'' wavevector modes propagate
subluminally, but the ``very large'' wavevector modes, i.e.,  with
$k$ near $k_0$, propagate superluminally.  As already mentioned, the
latter superluminal behavior is not essential for what we discuss in this
paper, it could be removed by considering a dispersion relation
like Unruh's \cite{Unruh95} in which the slope of the dispersion curve
asymptotes to zero as $k$ goes to infinity.

An important property of the action (\ref{action}), when generalized to
a complex scalar field,
is that it is invariant under constant
phase transformations of the field.
This leads to a conserved current
$j^{\mu}$.  The time component $j^{0}$, when integrated over a spatial
slice, serves as a conserved inner product when evaluated on solutions to the
equation of motion (\ref{eom}).  For the metric (\ref{metric}), the inner
product
takes the form
\begin{equation}
(F,G) = i \int dx\,  \Bigl(F^{*}
(\partial_{t} + v\partial_{x})G - G(\partial_{t} + v\partial_{x})
F^{*}\Bigr),
\label{inner}
\end{equation}
where $F(t,x)$ and $G(t,x)$ are solutions to (\ref{eom}).
Two classes of solutions to the field equation (\ref{eom}) are of interest.
The first are the positive free fall
frequency wavepackets.  They
can be written as a sum of solutions satisfying
\begin{equation}
(\partial_{t} + v \partial_{x})F(t,x) = -i \omega^{\prime}F(t,x)
\end{equation}
where $\omega^{\prime} >0$.
The second are the positive Killing frequency wavepackets.
These are a sum of solutions of the form
$e^{-i \omega t} f(x)$ where $\omega > 0$.
A positive free fall frequency wavepacket need not have a positive
norm under (\ref{inner}) in general, but does when $v(x) = constant$.
A positive Killing frequency wavepacket also need not have
positive norm in general, but does when $v(x) = 0$.

To quantize the field we assume that $\hat\phi(x)$ is a self-adjoint
operator solution to the field equation which satisfies the
canonical commutation relations.  We define for a normalized positive
free fall frequency solution $f(t,x)$ the
annihilation operator $a(f)$ by
\begin{equation}
a(f) \equiv (f,\hat\phi).
\end{equation}
We make a similar definition for the annihilation and creation
operators for a normalized
positive Killing frequency solution $g(t,x)$
(assuming it is also positive norm).   

\section{Computing the Particle Production}
\label{computing}

The standard method of computing the amount of particle production
in a given wavepacket is to propagate this wavepacket back in
time to the hypersurface where the quantum state boundary condition
is defined.  In our case we are interested in computing the particle
production in a late time, outgoing (right-moving), positive Killing
frequency wavepacket.  We assume that the state of the field is the
free fall vacuum, which is defined by
$a(p)|{\rm ff} \rangle =0$ for all positive
free fall frequency modes $p$ on the early time hypersurface.
Denoting our late time wavepacket by $\psi_{out}$, one may show that its
number expectation value in the free fall vacuum is 
\begin{equation}
N(\psi_{out}) = \langle {\rm ff} | a^{\dagger}(\psi_{out}) 
a(\psi_{out}) | {\rm ff} \rangle
= -(\psi_-,\psi_-)
\end{equation}
where $\psi_-$ is the negative free fall frequency part of $\psi_{out}$
after being propagated back to the early time hypersurface (see
\cite{CJ} for a detailed derivation).

Rather than solving the full equation of motion, (\ref{eom}), we shall
instead restrict ourselves to mode solutions of the form
\begin{equation}
\psi(t,x) = e^{-i \omega t} \phi(x).
\label{mode}
\end{equation}
Substituting into (\ref{eom}) (and setting $k_0 = 1$) produces 
the ordinary differential
equation (ODE)
\begin{equation}
\pm \phi^{(iv)}(x) + (1 - v^2(x))\phi^{\prime \prime}(x) + 2 v(x) (i \omega
- v^{\prime}(x)) \phi^{\prime}(x) - i \omega ( i \omega - v^{\prime}(x))
\phi(x) = 0
\label{ODE}
\end{equation}
where the $\pm$ refers to ${\hat F}_{\pm}$ respectively and we have used
a prime ($\prime$) to denote a derivative with respect to $x$. 
Restricting ourselves to mode solutions (\ref{mode})
therefore has the advantage
that we need only solve an ODE.

To determine the boundary conditions for (\ref{ODE})
it is necessary to study
wavepacket propagation in these models.
For the subluminal,
${\hat F}_{+}$, equation this has been discussed in great detail
in \cite{BMPS,CJ}.  The conclusion is that 
the late time, positive Killing frequency packet comes from
a pair of ingoing, short wavelength packets, located far outside
the black hole, and nothing\footnote{There could also
be an ingoing, long wavelength packet which arises from scattering
as the outgoing, positive Killing frequency packet is propagated
back in time.  For a slowly varying background spacetime we expect that
the amount of scattering is negligible, which is indeed what we find.}
else.  In particular, no part of the
wavepacket piles up against the horizon as with the ordinary wave
equation, and furthermore nothing comes from across the horizon.
This led \cite{CJ} to conclude that the boundary condition
for (\ref{ODE}) in the subluminal case is that the
solution decays across (and inside) the horizon\footnote{The
solution cannot vanish inside the horizon
because the coefficient of the highest derivative
term does not vanish, as it does with the ordinary
wave equation.}.

For the superluminal, $\m$, equation we refer the reader to 
\cite{Unruhpc,CJ971} for
the details of wavepacket propagation.  The conclusion
is that the late time, positive Killing frequency packet
comes from a pair of right-moving, short wavelength packets located far
$inside$ the horizon.  This is hardly surprising given that the $\m$
dispersion relation produces superluminal wave propagation.  The important
point concerning the boundary conditions though is that the only
packet ever outside the horizon is the late time, positive Killing
frequency packet\footnote{This is not quite correct.  Propagating the
outgoing, positive Killing frequency
packet back in time, a piece of it will scatter
into an early time, ingoing packet.  
We again expect that the amount of scattering is very
small for a slowly varying background spacetime, and in the 
leading order approximation discussed below we can ignore it.  This problem
can be dealt with by instead looking for mode solutions which
correspond to propagating positive free fall frequency packets
forward in time, which are necessary for computing the full quantum
state.  This will be discussed in subsection \ref{supstate}.}. 
This leads to the boundary conditions
for the superluminal ODE that for $x \gg 0$ the solution reduces to a single
mode with wavevector corresponding to the outgoing wavepacket.

Once we have solutions to the ODE's satisfying the boundary conditions
just discussed, we may easily extract the particle creation. 
In the subluminal case,
the solution at $x \gg 0$ (where $v(x) \approx constant$)
can be decomposed as
\begin{equation}
\phi(x) = \sum^{4}_{l=1} c_l (\omega) e^{i k_l(\omega) x},
\end{equation}
where $k_l (\omega)$ are the roots to the subluminal dispersion relation
(\ref{dispreln}).  From 
figure \ref{dispgraph} it is easy to see that two of these roots are
positive and the other two are
negative.
The late time wavepacket corresponds
to the small, positive wavevector ($k_{+s}$), and the early time, ingoing
wavepackets to the large, positive wavevector ($k_+$) and
the large, negative wavevector ($k_-$) respectively. 
The small negative wavevector ($k_{-s}$) corresponds to a long wavelength, 
ingoing wavepacket which will not be important in this leading
order calculation.  The number expectation
value for a mode of Killing frequency $\omega$ is then
\begin{equation}
N(\omega)=
\frac{|\omega'(k_{-})v_g(k_{-})c_{-}^2(\omega)|}
{|\omega'(k_{+s})v_g(k_{+s})c_{+s}^2(\omega)|}.
\label{No}
\end{equation}
The kinematic factors $v_g (k)$ and $\omega^{\prime} ( k)$ are the
group velocity as measured by a static observer and the frequency
as measured by a freely falling observer of a wavepacket narrowly
peaked
about wavevector $k$.

The superluminal equation can be 
handled in almost exactly the same manner.  Far
outside the horizon, where $v(x)$ is approximately constant and
satisfies $0>v(x)>-1$, it is again easy to see from figure \ref{dispgraph}
that there is one positive wavevector root ($k_{+s}$)
of the dispersion relation 
(\ref{dispreln}) (with the plus sign) and one negative 
wavevector root ($k_{-s}$).
Our boundary conditions dictate that the
solution at $x \gg 0$ is\footnote{Note that we use the
same notation for denoting the wavevectors for both the subluminal and
superluminal
cases; however, the actual values of the wavevectors for a given
$\omega$ differ between the two cases.}
\begin{equation}
c_{+s} e^{i k_{+s} (\omega) x}. 
\end{equation}
To avoid dealing with the singularity, we shall also
assume that $v(x)$ becomes constant behind the horizon.  When the slope
of the straight line in
figure \ref{dispgraph} is larger than one, it is easy to
see that there is one positive 
wavevector root ($k_+$) to the dispersion relation
(\ref{dispreln}) (with the positive sign) and
three negative wavevector roots which we denote as $k_{-s}$, $k_{-m}$, and
$k_-$ in order of increasing magnitude ($s$ denoting $small$ and $m$ denoting
$middle$).  As 
we shall see below, only the large positive
and large negative wavevector solutions will contribute to the solution. 
It follows that
the solution in this region is of the form
\begin{equation}
c_{+} e^{i k_{+} (\omega) x} +
c_{-} e^{i k_{-} (\omega) x}. 
\end{equation}
The number expectation value in this case again becomes (\ref{No}), with
the kinematic factors appropriate for the superluminal equation, $\m$.

\section{Approximate solutions to the subluminal equation}
\label{sub}

The methods applied to find approximate solutions to ODE (\ref{ODE}) are the
same as those used in solving the Schrodinger equation for a tunneling
potential.  For the general potential one may find approximate
solutions to the Schrodinger equation by the WKB method \cite{Schiff};
however, about a classical turning point (i.e., where the kinetic energy
vanishes) the WKB approximation breaks down.  
Approximate solutions can nevertheless be obtained by expanding the potential
$V(x)$ in the full Schrodinger equation about the classical turning
point $x_{tp}$.
Solutions to the resulting equation are straightforward to find and
are valid in the region
\begin{equation}
|x - x_{tp}|^{n-1} \ll |n! V^{\prime}(x_{tp})/V^{(n)}(x_{tp})|
\end{equation}
if $V^{\prime \prime}(x_{tp}), \cdots, V^{(n-1)}(x_{tp})$ all vanish.
 
If the WKB solutions are valid in regions on either side of the
classical turning point which overlap with the region of validity of the 
solution to the linearized potential
equation, then we can obtain approximate solutions
to the full
Schrodinger equation over the entire range of $x$.  We now apply this
method to ODE (\ref{ODE}) for the subluminal equation, $\p$.  We begin by
finding the approximate solution about the horizon relevant for
particle creation in subsection \ref{subapprox}, and then match
this solution to WKB solutions outside the horizon in subsection 
\ref{subwkb}.  From this solution we then compute the amount of 
particle creation in subsection \ref{spectrum}.  Finally we show in
subsection \ref{substate}
that this method can be extended to computing the complete out-state
to leading order in $k_0$, i.e., compared to just computing number
expectation values.

\subsection{Approximate solutions about the horizon}
\label{subapprox}

\begin{figure}[hbt]
\centerline{
\psfig{figure=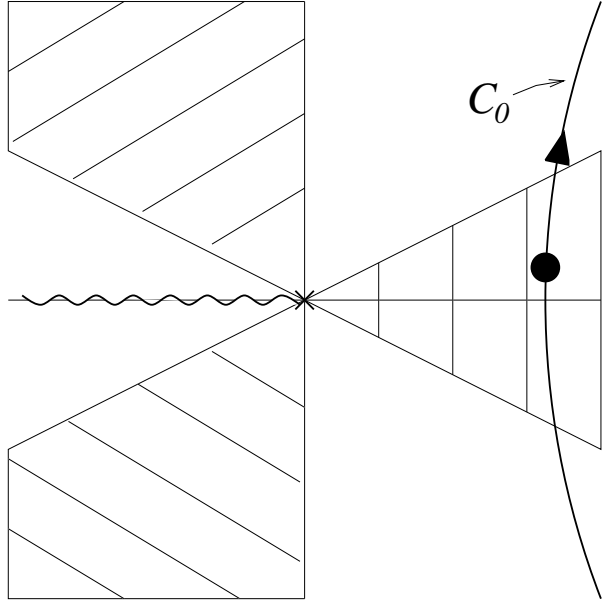,angle=-90,height=8cm}}
\caption{\small Diagram of the steepest descent contour $C_{0}$.
The unmarked regions are directions in which the contour must
asymptote for the integral to converge.  The $\times$'s are
singularities of the integrand and the wavy line is a branch cut.}
\label{contour1}
\end{figure}

\begin{figure}[hbt]
\centerline{
\psfig{figure=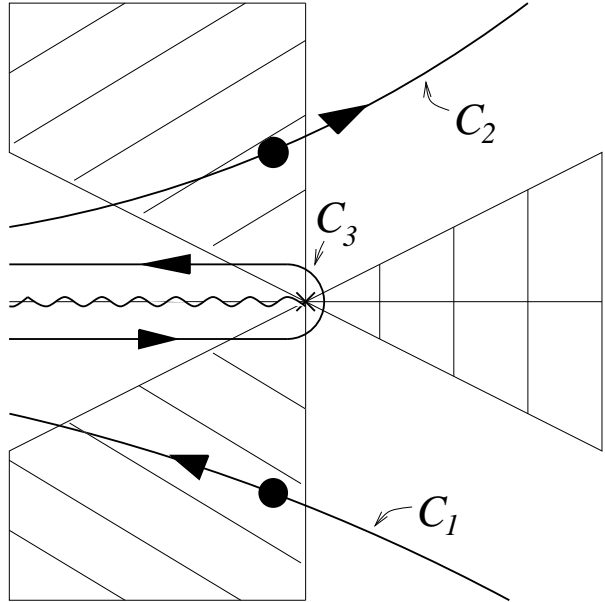,angle=-90,height=8cm}}
\caption{\small Diagram of the steepest descent contours $C_{1}$,
$C_{2}$, and $C_{3}$.  $C_{1}$ and $C_2$ pass through the
saddle points 
$s_+$ and $s_-$ respectively.
The unmarked regions are directions in which the contour must
asymptote for the integral to converge.  The $\times$'s are
singularities of the integrand and the wavy line is a branch cut.}
\label{contour2}
\end{figure}

In words the calculations we present in this subsection are
as follows.  We
find approximate solutions to ODE (\ref{ODE}) about the horizon 
by linearizing the $x$-dependent coefficients in the equation
and solving the resulting
equation by the method of Laplace
transforms.  These solutions are given by contour integrals in the
complex $s$-plane (where $s$ is the Laplace transform variable).  The
contour $C_0$ (see figure \ref{contour1})
corresponding to the $x$-space solution inside the horizon
is chosen so that the boundary conditions described in section \ref{computing}
are satisfied, i.e., that the solution decays inside the horizon, as these
are the boundary conditions relevant for particle creation.  The 
$x$-space solution outside the horizon must then
arise from a contour that is deformable to $C_0$.  This contour is
broken up into three separate contours:  $C_1, C_2$, and $C_3$ (see figure
\ref{contour2}).  By comparing the solutions corresponding to these contours
to the WKB solutions (computed in subsection \ref{subwkb}) of ODE (\ref{ODE}),
we show that $C_3$ corresponds to the
late time, outgoing Hawking particle and that
$C_1$ and $C_2$ 
correspond to the ingoing, large positive
and negative wavevector packets respectively from which the outgoing
Hawking particle arises.  Once this has been done it is a simple matter
to compute the particle creation as discussed in section \ref{computing}.

We first linearize $v(x)$ and $v^{\prime}(x)$ about the horizon as
\begin{eqnarray}
v(x) \approx -1 + \kappa x \label{vexpand} \\
v^{\prime}(x) \approx \kappa + \kappa_{1}^2 x \label{vpexpand}
\end{eqnarray}
where $\kappa$ is the surface gravity of the black hole described
by the metric (\ref{metric}) and $\kappa_1$ is a higher order correction to
$v(x)$.
Substituting into (\ref{ODE}) and keeping only linear terms in $x$ yields
\begin{equation}
\phi^{(iv)}(x) + 2 \kappa x \phi^{\prime \prime}(x) + 2(-(i \omega - \kappa)
+(\kappa ( i \omega - \kappa) + \kappa_{1}^2)x) \phi^{\prime}(x) -
i \omega (i \omega - \kappa - \kappa_{1}^2 x)\phi(x) = 0.
\label{ODEtemp}
\end{equation} 
Validity of this equation requires that $|\kappa x| \ll 1$ and
$|\kappa_{1}^2 x/\kappa| \ll 1$.  To leading order
we may therefore further simplify the equation as
\begin{equation}
\phi^{(iv)}(x) + 2 \kappa x \phi^{\prime \prime}(x) - 2(i \omega - \kappa)
\phi^{\prime}(x) -
i \omega (i \omega - \kappa)\phi(x) = 0.
\label{ODE1}
\end{equation} 
This is the equation we shall use in this paper to find approximate solutions
about the horizon; however, to compute correction terms to the flux,
we must keep at least all linear terms in $x$ and possibly
even higher order terms in $x$.  We shall discuss this further in the 
conclusions section.

We use the method of Laplace transforms, \cite{Bleis,Wong}, to solve
(\ref{ODE1}).  Writing the solution as a Laplace transform,
\begin{equation}
\phi(x) = \int_{C} ds \, e^{sx} \phit(s),
\label{laplace}
\end{equation}
(where $C$ is the contour of integration) and substituting into 
(\ref{ODE1}) yields the $s$-space\footnote{We 
have dropped a boundary term to obtain (\ref{sODE}).  This term
will vanish by our choice of contours  below.}
ODE 
\begin{equation}
\partial_s \left( \ln(s^2 
\phit(s)) \right) = \frac{s^4 -
2 (i \omega - \kappa) s - i \omega ( i \omega - \kappa)} {2 \kappa
s^2}.
\label{sODE}
\end{equation}
Equation (\ref{sODE}) is easily solved as
\begin{equation}
\phit(s) = s^{-1 - i \omega/\kappa} 
\exp \left( \frac{1}{2 \kappa} \left(\frac{s^3}{3} + \frac{i \omega (i \omega
- \kappa)}{s} 
\right) \right)
\label{ssoln}
\end{equation}

To obtain the $x$-space solution, we substitute
$\phit(s)$ into (\ref{laplace}) and integrate.  The choice of contour
$C$ over which we integrate is dictated by the boundary conditions discussed
in section \ref{computing}, i.e., we want $\phi(x)$ to decay inside the horizon.
Before finding the appropriate contour to produce this behavior let's
first understand the generic properties that the contour must satisfy.
Specifically, note that $\phit(s)$ is dominated at large $|s|$
by the $\exp(s^3/(6 \kappa))$ term, and therefore for the
integral to converge (assuming the contour $C$ runs to infinity, which 
it need not) the contour must asymptote to a region where $Re(s^3) < 0$
since $\kappa$ is real and positive.  Writing $s = r e^{i \theta}$, this
implies that the contour must asymptote to any of the three regions
\begin{eqnarray}
Region \, 1 \,&  \leftrightarrow & \, 
\frac{\pi}{6} < \theta < \frac{\pi}{2} \nonumber \\
Region \, 2 \, & \leftrightarrow & \, \frac{5 \pi}{6} < \theta < \frac{7 \pi}{6}
\nonumber \\
Region \, 3 \, & \leftrightarrow & \, 
\frac{3 \pi}{2} < \theta < \frac{11 \pi}{6}.
\label{regions1}
\end{eqnarray}
In figure 
\ref{contour1} these appear as the unmarked regions.  

To evaluate the contour integral (\ref{laplace}) we first consider
the $x<0$ case.  We must choose a contour that yields a solution 
that decays with decreasing $x$ inside the horizon, and that
asymptotes to any of the three regions (\ref{regions1}).  To find
this solution we approximate the contour integral (\ref{laplace}) by
the method of steepest descents, \cite{Wong}.

First rewrite the contour integral as
\begin{equation}
\phi(x) = \int_{C} ds \, g(s) e^{x f(s)}
\label{contint}
\end{equation}
where
\begin{equation}
g(s) = s^{-1 - i \omega/\kappa} 
\label{gofs}
\end{equation}
and
\begin{equation}
f(s) = s + \frac{1}{2 \kappa x} \left( \frac{s^3}{3} + 
\frac{i \omega (i \omega - \kappa)}{s} \right).
\label{fofs}
\end{equation}
To evaluate (\ref{contint}) by steepest descents we first
locate the saddle
points of $f(s)$.
These are given by the roots of
$df(s)/ds = 0$, \cite{Wong}, which in this case are approximated
by
\begin{equation}
s_{\pm} \approx \pm \sqrt{- 2 \kappa x}
\label{spmroots}
\end{equation}
(since we are in a region where $\omega, \kappa \ll |\kappa x|$).
The contours of steepest descent through these saddle points are
given by $Im(f(s) - f(s_{\pm})) = 0$ and $Re(x(f(s) - f(s_{\pm}))) < 0$.
Using this one may show that the direction of the steepest descent
contours through $s_{+}$ and $s_{-}$ are $-\pi/2$ to $\pi/2$ and
$0$ to $\pi$ respectively.  It is not hard to show that the steepest
descent contour, $C_0$, through $s_{+}$ asymptotes to regions 1 and 3 as
shown in figure \ref{contour1}.  The contour integral in this case is now
obtained by the standard formula
\begin{equation}
\phi_{0}(x) \approx g(s_+) \sqrt{\frac{2 \pi}{|x f^{\prime \prime}(s_+)|}}
e^{x f(x_+) + i \alpha_+}
\label{phisaddle}
\end{equation}
where $\alpha_+=\pi/2$ if we traverse the contour in the direction indicated
in figure \ref{contour1}.  To lowest order in $\omega$ and $\kappa$ this
reduces to
\begin{equation}
\phi_0(x) \approx - \sqrt{2 \pi \kappa} \, (-2 \kappa x)^{-3/4 -
i \omega/(2 \kappa)}
\exp \left(-\frac{2}{3} \sqrt{2 \kappa} |x|^{3/2} \right).
\label{phip}
\end{equation}
We immediately see that this is exponentially decaying with decreasing
$x$ (recall that $x<0$).  The contour through $s_-$ produces an exponentially
growing solution, hence our desired solution is given by the contour
$C_0$.  Finally, note that $g(s)$ is singular at $s=0$, and
that we must choose a branch cut from this point.  We choose the branch cut
to run along the negative real $s$-axis.

At this point the reader may be wondering about the validity of the
approximations made so far.  The steepest descents method requires
that $|x| \gg 1$ while validity of the approximate ODE (\ref{ODE1})
requires that $|\kappa x| \ll 1$. 
As long as $\kappa \ll 1$, there is always a wide range of $x$ values
satisfying both conditions, i.e., $1 \ll |x| \ll 1/\kappa$.  Such 
$\kappa$ correspond to black
hole temperatures $T_H \ll 1$ (or $T_H \ll k_0$ if we restore $k_0$).  For
example, if the Planck length is also one in these units (i.e., $k_0=
1=1/l_{Pl}$), and $\kappa$ is the surface gravity of a solar mass black
hole, the inequality on $x$ becomes $1 \ll |x| \ll 10^{38}$.  Clearly
there is no problem in satisfying this inequality.
It is convenient to keep these numbers in
mind for later approximations.

Now we turn to evaluating $\phi(x)$ for $x > 0$.  In principle we must
evaluate the contour integral (\ref{contint}) over the same contour as
in the $x < 0$ case, i.e., $C_0$.  However, by Cauchy's theorem we may
deform the contour (keeping the endpoints fixed) through 
any region in which the integrand is analytic
to a new contour, hopefully one where the integral is easier to evaluate.
In particular we may deform the contour so that it runs through any nearby
saddle points so that we may again approximate the integral by the method
of steepest descents.
In fact most of the work for these saddle points has already been
done.  They are still given by
(\ref{spmroots}) except that now $x > 0$ and therefore are both
imaginary (to leading
order), i.e., $s_{\pm} \approx \pm i \sqrt{2 \kappa x}$.
The direction of the steepest descent contours through
$s_-$ and $s_{+}$ are now given by $7 \pi/4$ to $3 \pi/4$ and
$5 \pi/4$ to $\pi/4$ respectively.  From this one can easily see that
the steepest descent contour, $C_1$, through $s_-$ asymptotes to
regions 2 and 3 (\ref{regions1}) and
the steepest descent contour, $C_2$, through
$s_+$ asymptotes to regions 1 and 2 (\ref{regions1}), as shown
in figure \ref{contour2}.
Evaluating the leading order contributions to these contour integrals
as before results in
\begin{eqnarray}
\phi_1(x) & \approx & e^{i 5 \pi/4} e^{- \pi \omega/(2 \kappa)} \sqrt{2 \pi
\kappa} \, (2 \kappa x )^{-3/4 - i \omega/(2 \kappa)} \exp \left( -i
\frac{2}{3} \sqrt{2 \kappa} x^{3/2} \right) \label{phi1} \\
\phi_2(x) & \approx & e^{-i \pi/4} e^{ \pi \omega/(2 \kappa)} \sqrt{2 \pi
\kappa} \, (2 \kappa x )^{-3/4 - i \omega/(2 \kappa)} \exp \left( i
\frac{2}{3} \sqrt{2 \kappa} x^{3/2} \right) \label{phi2}
\end{eqnarray}
where we have chosen the directions of the contours as depicted in figure
\ref{contour2}.

The contour $C_1 + C_2$ is not by itself deformable to $C_0$, but if
we add in the contour $C_3$ (see figure \ref{contour2})
which asymptotes to region 2 on either
side of the branch cut, then $C_1 + C_2 + C_3$ is deformable to
$C_0$.  To evaluate the contour integral over $C_3$, first define
the new integration variable $e^{-i \pi} t := s \, x$.  This produces
\begin{equation}
\phi_{3}(x) = x^{i \omega/\kappa} \int_{{\bar C}_3} dt 
(-t)^{-1-i \omega/\kappa} \exp \left(-t + \left( -\frac{t^3}{6 \kappa x^3}
- \frac{i \omega ( i \omega - \kappa)x}{2 \kappa t} \right) \right)
\label{phi3cont}
\end{equation}
where the new contour ${\bar C}_3$ runs from infinity just above the 
postive real
axis, counter-clockwise about the origin, and back to infinity just
below the positive real axis.  Ignoring the $t^3$ and $t^{-1}$ terms in the
exponent for the moment, we note that the remainder is just a gamma
function, i.e., using the integral representation
\cite{GradRyz},
\begin{equation}
\Gamma(\nu) = - \frac{1}{i 2 sin(\pi \nu)} \int_{\bar{C}_3} dt \, 
(-t)^{-1 + \nu}
e^{-t}
\label{gamma}
\end{equation}
we arrive at
\begin{equation}
\phi_3(x) \approx -2 \sinh(\pi \omega/\kappa) \, \Gamma(-i\omega/\kappa) \,
x^{i \omega/\kappa}.
\label{phi3}
\end{equation}

To see when this approximation holds, expand
\begin{equation}
\exp \left(- \frac{t^3}{6 \kappa x^3} \right) \approx 1 - \frac{t^3}
{6 \kappa x^3} +  O \left( \frac{t^6}{\kappa^2 x^6} \right).
\end{equation}
Evaluating the integral (\ref{phi3cont}) with the $t^3$ 
term produces the correction term to $\phi_3(x)$,
\begin{equation}
\delta \phi_3(x) = \frac{\sinh(\pi \omega/\kappa)}{3 \kappa x^3} \,
\Gamma(-i \omega/\kappa) x^{i \omega/\kappa}.
\end{equation}
Using the identity $\Gamma(z+1) = z \Gamma(z)$ one may show that 
$|\delta \phi_3(x)/\phi_3(x)| \ll 1$ holds if 
\begin{equation}
1 \ll | \kappa x^3|.
\label{cond1}
\end{equation}
If we expand $\exp (-i \omega (i \omega - \kappa) x/(2 \kappa t))$ in the
same way and evaluate the leading order correction term to
$\phi_3(x)$ as before, we find that we need 
\begin{equation}
|\omega x| \ll 1.
\label{cond2}
\end{equation}
When these conditions hold, $\phi_3(x)$ is well approximated by (\ref{phi3})
(in the asymptotic expansion sense).

To summarize, we have found an approximate solution $\phi(x)$ to the mode
equation (\ref{ODE}) satisfying the boundary conditions that it
decay inside the horizon.  Just outside the horizon the solution is given
by $\phi(x) =
\phi_1(x) + \phi_2(x) + \phi_3(x)$ (see (\ref{phi1}), (\ref{phi2}), and
(\ref{phi3}) respectively).  We now propagate this
solution out to a region where $v(x)$ is essentially constant
by patching onto WKB solutions which are valid outside the
horizon.  Knowing the solution in the constant
$v(x)$ region will then allow us to easily extract the particle
flux.

\subsection{WKB solutions}
\label{subwkb}

To find approximate solutions to the mode equation (\ref{ODE}) by the
WKB method,
assume a solution of the form
\begin{equation}
\phi(x) = e^{i \int dx \, k(x)}
\end{equation}
where the wavevector $k(x)$ is an unknown function of $x$.  Substitution
yields 
\begin{eqnarray}
& & k^4 - (1 - v^2) k^2 - 2 v \omega k + \omega^2 = \nonumber \\
& &  i \frac{d}{dx}(2 k^3
-(1-v^2) k - \omega v) + (4 k k^{\prime \prime} + 3 (k^{\prime})^2) -
i k^{\prime \prime \prime}
\label{WKB1}
\end{eqnarray}
where we denote derivatives with respect to $x$ by primes $(\prime)$.
If $v(x)$ is a slowly varying function of $x$, then we expect $k(x)$ also
to be slowly varying.  We therefore try to solve for $k$ perturbatively
in derivatives of $v(x)$.  More precisely, let $x \rightarrow \alpha x$
(we will take $\alpha=1$ at the end), then (\ref{WKB1}) becomes
\begin{eqnarray}
& & k^4 - (1 - v^2) k^2 - 2 v \omega k + \omega^2 =  \nonumber \\
& & \frac{i}{\alpha} \frac{d}{dx}(2 k^3
-(1-v^2) k - \omega v) + \frac{1}{\alpha^2} (4 k k^{\prime \prime} 
+ 3 (k^{\prime})^2) -
\frac{i}{\alpha^3} k^{\prime \prime \prime}.
\label{WKB2}
\end{eqnarray}
In this equation we see on the right-hand-side that derivatives
of $k(x)$ and $v(x)$ with respect to $x$ are suppressed by powers of 
$1/\alpha$.

Now assume that $k(x)$ may be expanded in inverse powers of $\alpha$ as
\begin{equation}
k(x) = k^{(0)}(x) + \frac{1}{\alpha} k^{(1)}(x) + \cdots.
\end{equation}
Substituting into (\ref{WKB2}) and demanding that the coefficients of each
power of $1/\alpha$ separately vanish produces an infinite set of equations,
the lowest orders being
\begin{eqnarray}
& & (k^{(0)})^4 - (1 - v^2) (k^{(0)})^2 - 2 v \omega k + \omega^2 = 0 
\label{k0eqn} \\
& & k^{(1)} = \frac{i}{2}
\frac{d}{dx} \ln ( 2 (k^{(0)})^3 - ( 1-v^2) k^{(0)} - v \omega).
\label{k1eqn}
\end{eqnarray}
Although the leading order equation for $k^{(0)}$, (\ref{k0eqn}), can 
be solved exactly
producing a set of four wavevectors, the expressions are quite unwieldy.
Fortunately, since we are mainly interested in Killing frequencies $\omega$
satisfying $\omega \ll 1$, we only need find approximate wavevector roots
to this equation.  Once these roots are known, the $1/\alpha$ corrections to
them can be found by substituting the respective wavevector root $k^{(0)}$
into (\ref{k1eqn}) and solving for $k^{(1)}$.  These computations produce the
wavevectors
\begin{eqnarray}
k_{\pm} & = & \pm \sqrt{1-v^2} +\frac{\omega \, v}{1 - v^2} + 
i \frac{3}{4} \frac{d}{dx} \ln(1-v^2) +
{\cal O}(\omega^2) \label{pmroots} \\
k_{+s} & = & \frac{\omega}{1+v} + {\cal O}(\omega^3) \label{sproot} \\
k_{-s} & = & - \frac{\omega}{1-v} + {\cal O}(\omega^3) \label{snroot},
\end{eqnarray}
where we have set $\alpha =1$.  The corresponding WKB solutions are
\begin{eqnarray}
\phi_{\pm}(x) & \approx & (1-v(x)^2)^{-3/4} e^{\pm i \int dx \, \sqrt{1-v(x)^2}}e^{i \omega \int dx \, v(x)/(1 - v^2(x))}
\label{pmWKB} \\
\phi_{+s}(x) & \approx &  e^{i \omega \int dx/(1+v(x))} \label{spWKB} \\
\phi_{-s}(x) & \approx & e^{-i \omega \int dx/(1 - v(x))} \label{snWKB}.
\end{eqnarray} 
The condition of validity for these approximate solutions is that
$|k^{(1)}(x)/k^{(0)}(x)| \ll 1$.  For the $k_{\pm}$ wavevectors this ratio
is
\begin{equation}
\left| \frac{k^{(1)}_{\pm}(x)}{k^{(0)}_{\pm}(x)} \right| \approx
\frac{3}{2} \left| \frac{v(x) v^{\prime}(x)}{(1-v^2(x))^{3/2}} \right|.
\label{WKBvalid}
\end{equation}
We are interested in $v(x)$'s containing black holes.  The horizon of
a black hole in these units is located at $v(x_h)=-1$, therefore the
right-hand-side of (\ref{WKBvalid}) clearly becomes arbitrarily large
as we approach the horizon (assuming that $v^{\prime}(x) \neq 0$ which
are the only cases we consider here).  It follows that the WKB approximation
will break down around the horizon.  Far from the horizon (and outside
the black hole) $v(x)$ asymptotes
to a constant $-1 < v_0 < 0$ and $v^{\prime}(x)$ goes to zero, therefore
the WKB approximation will be valid.  
For the $k_{+s}$ mode a ratio similar to
(\ref{WKBvalid}) holds, and therefore the WKB approximation again fails
around the horizon, but is valid far outside of it.  To compute
this ratio, we must compute $k_{+s}$ to order $O(\omega^3)$, however
since we will not need this later we do not give the explicit
expressions here.

\subsection{The spectrum}
\label{spectrum}

We now have all the ingredients necessary to compute the leading order spectrum
of black hole radiation.  First note that validity of the
WKB approximation (\ref{WKBvalid}) requires that
$1 \ll \sqrt{\kappa x^3}$.  Furthermore the
approximate solution from the Laplace
transform method is valid when $1 \ll |x| \ll 1/\kappa$.
Since we are considering cases where $\kappa \ll 1$, there
is always a region where both the WKB and Laplace transform solutions
are valid.

If we evaluate the integrals appearing in the
WKB solutions of (\ref{pmWKB}), (\ref{spWKB}) and, (\ref{snWKB}) 
respectively in a
region just outside the horizon where $v(x)$ is
given by the linearized expression (\ref{vexpand}), we find
that the solution outside the horizon obtained
from the Laplace transform method
can be expressed as
\begin{eqnarray}
\phi(x) & = & e^{-i \pi/4} \, \sqrt{2 \pi \kappa} \,
(-i \,e^{- \pi \omega/(2 \kappa)}
\, \phi_{-}(x)
+ e^{\pi \omega/(2 \kappa)} 
\, \phi_{+}(x)) \nonumber \\
& - & e^{- \pi \omega/\kappa} ( e^{2 \pi \omega/\kappa} -1) \Gamma(- i
\omega/\kappa) \, \phi_{+s}(x).
\label{wkbform}
\end{eqnarray}
With $\phi(x)$ decomposed in terms of the WKB solutions, we 
are allowed to evaluate it at large
$x$, i.e., where $v(x)$ is essentially constant.  In this region the
WKB solutions reduce to simple modes (up to multiplicative constants)
and we need only extract the coefficients of these modes in order
to compute the particle creation rate as given in (\ref{No}).
A simple computation yields
\begin{equation}
N(\omega) = \frac{1}{e^{2 \pi \omega/\kappa} - 1},
\label{flux}
\end{equation}
exactly a thermal spectrum at the Hawking temperature $T_{H} = \kappa/(2
\pi)$.

As a check on our results,
recall that invariance of the action (\ref{action}) under constant phase 
transformations of the field leads to the existence of a conserved current
$j^{\mu}$.  When this current is evaluated on fixed Killing frequency
mode solutions, the time component $j^t$ is manifestly
time independent and the conservation law reduces to $\partial_x j^x =0$,
i.e., $j^x$ is constant.  The exact form of $j^x$ is complicated, but
when evaluated on a mode of the form $c(\omega)
 \exp (- i \omega t + i k(\omega) x)$ in
a region where $v(x)$ is constant, it reduces to 
\begin{equation}
j^x(k(\omega)) = \omega^{\prime}(k(\omega)) v_g (k(\omega)) |c(k(\omega))|^2.
\end{equation}
For the solution of interest to us, i.e., the one that decays inside
the horizon, the spatial part of the current must vanish everywhere.
Using the solution given by (\ref{wkbform}), it is easy to show that
this is indeed the case.   

We have made a number of approximations in order to compute the leading
order flux given by (\ref{flux}), we would now like to collect them
to find out what restrictions they place on the allowed parameter
ranges.  First recall that we have restricted $\kappa$ to values
$\kappa \ll 1$ (in units of $k_0 =1$).   Physically this says
that we only expect to get a thermal spectrum of radiation when
the black hole is large, and therefore has a small temperature.

The range of validity of the Laplace transform solutions in the
spatial variable $x$ is also restricted.  We have already seen that
we need $|x| \gg 1$ for the steepest descents approximation to
hold for the various contour integrals and
$|\kappa x| \ll 1$ for the approximate ODE (\ref{ODE1}) to be valid.
Closer investigation of the correction terms to the WKB solutions
and the Laplace transform solutions shows that we need
\begin{equation}
1/\kappa^{1/3} \ll |x| \ll 1/\kappa^{3/5}.
\label{xineq}
\end{equation}
This inequality means that the matching of the WKB and Laplace transform
solutions can be done anywhere in this range.
To derive these inequalities, note that the WKB and Laplace transform
solutions cannot be matched to arbitrary order because in fact they
solve different equations.  The Laplace transform solutions were obtained
by finding approximate solutions to the linearized $v(x)$
ODE (\ref{ODE1}), while
the WKB solutions were obtained by finding approximate solutions 
to the full ODE.  By computing the first set of correction terms to both the
WKB and Laplace transform solutions that differ, i.e., do not match, and
demanding that they are small (compared to the leading order
terms), we derive the above inequality.

These correction terms also restrict the range of $\omega$.  Since the
corrections
are $x$ dependent, by matching
the WKB and Laplace transform solutions about an appropriate $x$ satisfying
(\ref{xineq}) we can minimize the difference between the solutions.
To carry this out we compute the leading order correction terms for
the WKB and Laplace transform solutions that differ and sum their absolute
values.  As a specific case, the leading order relative difference 
between the WKB solution corresponding to the wavevector $k_+$
and the Laplace transform solution corresponding to the contour $C_2$
is
\begin{equation}
\sigma \approx \left( \frac{2}{\pi^2} \right)^{1/4} \frac{\omega}{\kappa}
\frac{\sqrt{\kappa}}{(2 \kappa x)^{3/4}} + \frac{1}{40} 
\frac{(2 \kappa x)^{5/2}}{\kappa}
\end{equation}
in the limit of $\omega \gg \kappa$.  Minimizing this with respect
to $x$ we find that 
$\kappa x_{min} \approx (9 \kappa \omega^2/(4 \pi))^{2/13}$ 
and 
\begin{equation}
\sigma \approx \frac{13}{120} \left( \frac{2}{\pi^2} \right)^{5/26}
12^{10/13} \, \frac{\omega^{10/13}}{\kappa^{8/13}}.
\end{equation}
Demanding that $\sigma \ll 1$ we arrive at the constraint
\begin{equation}
\omega \ll \frac{1}{12} \left( \frac{\pi^2}{2} \right)^{1/4} \left(
\frac{120}{13} \right)^{13/10} \kappa^{4/5}.
\end{equation}
A similar computation can also be carried out for the WKB and Laplace
transform solutions corresponding to the $k_-$ and
$k_{+s}$ wavevector roots.
In the limit of $\omega \gg \kappa$ a weaker constraint than
above is obtained from the $k_{+s}$ root, and the same
constraint as above is obtained for the $k_-$ wavevector.
In the same manner as above a lower bound on the allowed range
of $\omega$ can be obtained.  From the $k_+$ wavevector we
find that we need
\begin{equation}
\omega \gg \frac{13}{6 \pi} \left( \frac{2}{\pi} \right)^{5/13} 
\frac{1}{20^{3/13}} \kappa^{15/13}.
\end{equation}
It follows that the range of validity of the analytical results presented
above are bounded both above and below in the parameter $\omega$.

\subsection{Computing the quantum state}
\label{substate}

We have so far computed the outgoing flux of particles from a black
hole for the subluminal equation of motion (\ref{eom}) (with the plus
sign).
There is, however, much more information contained in the quantum
state than just number expectation values, for instance, correlations.
It is therefore of interest to compute the 
full quantum state in this modified theory and compare it to
the state arising with the ordinary wave equation, as already
computed by Wald \cite{Wald75}.

To be a bit more precise, we are not actually ``computing'' the
quantum state because we already know what it is, i.e.,
we have assumed that it is the free fall vacuum.  What we are going
to do is re-express this state in terms of a vacuum state defined
by late time observers.  We define the out-Hilbert space
as the tensor product of Hilbert spaces on either side of the
horizon.  Outside the horizon we use Killing frequency to define
the Hilbert space, as we have done so far.  Inside the horizon
we do the following.  We take a $v(x)$ that asymptotes to a
constant (less than $-1$), then we define our Hilbert space in this
region using free fall frequency.  If we are only interested in
the observations made by the 
outside observer then we would trace over the inside degrees
of freedom, in which case the Hilbert space we use inside the
horizon is irrelevant.  For the computations here though, the choice
we have made is the simplest.

Our method of computing the state is very similar to the
techniques employed by Wald \cite{Wald75}.  In the time
dependent picture we would take an ingoing, positive free
fall frequency wavepacket and evolve it from the hypersurface
where the free fall vacuum is defined to the hypersurface
where the out vacuum is defined.  If this packet is sufficiently
peaked in its wavevector, we may follow it on the dispersion
relation as discussed in detail in \cite{CJ}.  It 
is not hard to see that this
packet will propagate toward the horizon and scatter (mode convert),
the reflected piece propagates (forward in time) away from the horizon
to the region where $v(x)$ is essentially constant, and the transmitted
packet propagates deeper inside the horizon to where $v(x)$ is
essentially constant.  The late time packet outside the horizon corresponds
to the Hawking particle, and the late time packet inside the horizon
corresponds to the partner.
If the initially positive free fall frequency
packet $\psi_{+{\rm ff}}$ contains only positive Killing frequencies, the 
final packet
$\psi_{+{\rm out}}$
outside the horizon will contain only positive Killing frequencies 
and the final packet $\psi_{-{\rm in}}$ inside the horizon will contain only
negative free fall frequencies.  The annihilation operator associated
with $\psi_{+{\rm ff}}$, i.e., 
$a(\psi_{+{\rm ff}}) := (\psi_{+{\rm ff}},\hat{\phi})$,
annihilates the free fall vacuum.  Using the time independence of the
inner product, we therefore derive the equation
\begin{equation}
(a(\psi_{+{\rm out}}) - a^{\dagger}(\psi^{*}_{-{\rm in}}))|{\rm ff} \rangle = 0.
\label{stateeqn}
\end{equation}

Similarly, if the initially positive free fall frequency
packet contains only negative Killing frequencies, the final packet
outside the horizon will contain only negative Killing frequencies
and the final packet inside the horizon will contain only
positive free fall frequencies.  Using this, a relation similar
to (\ref{stateeqn}) can be derived.  As shown by Wald \cite{Wald75},
given a complete set of relations like (\ref{stateeqn}) (constructed
by taking a
complete set of ingoing, positive free fall frequency packets),
we can re-express the free fall vacuum in terms of
the out vacuum.

As we have done thus far, we shall actually use mode solutions instead
of wavepackets.
We derive the mode solutions which, when appropriately summed, produce
the time dependent wavepacket solutions just discussed.  We have 
already derived one mode solution in (\ref{wkbform}), although
it is not of the form that we want.  Rather it
decays inside the horizon and is a superposition of plane waves
with wavevectors $k_{+s}$, $k_+$, and $k_-$ far outside the 
horizon (where $v(x)$ is essentially constant).
In the time dependent
picture it corresponds to propagating a pair of 
ingoing, positive and negative free fall frequency wavepackets forward
in time, with just the right relative weights so that the entire
packet completely
mode converts around the horizon, turns around, and propagates 
out to the constant $v(x)$ region.  To obtain the mode solutions
that we want, we shall construct another mode solution below
which is a superposition of plane waves with wavevectors
$k_+$ and $k_-$ far outside the horizon (where $v(x)$ is
essentially constant), whereas inside the horizon (where
$v(x)$ is again essentially constant) it reduces to a plane
wave with wavevector $k_-$.  
In the time dependent
picture this mode solution corresponds to propagating a pair of 
ingoing, positive and negative free fall frequency wavepackets forward
in time, with just the right relative weights (although
different than above) so that the entire
packet propagates across the horizon and converts into just
a negative free fall frequency packet.
By adding these two mode solutions, call them the n-modes since they
are the relevant ones for computing number expectation values, with 
the correct relative
coefficient, we can eliminate the $k_-$ ($k_+$)
mode outside
the horizon.  These are the mode solutions we want, call them the
s-modes since they are the relevant ones for computing the state,
because they
correspond to propagating an ingoing,
positive (negative) free fall frequency packet forward
in time which splits around
the horizon into a pair of wavepackets, one propagates back away from
the horizon, and the other falls inside the black hole.

\begin{figure}[hbt]
\centerline{
\psfig{figure=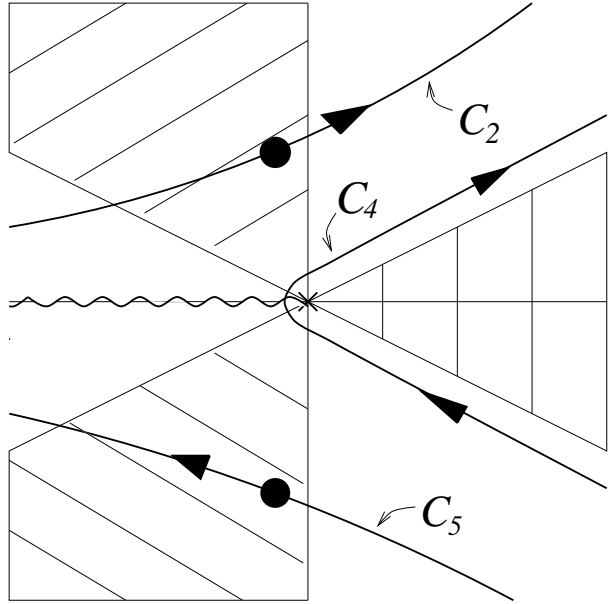,angle=-90,height=8cm}}
\caption{\small Diagram of the steepest descent contours $C_{2}$,
$C_{4}$, and $C_{5}$.  $C_{2}$ and $C_5$ pass through the
saddle points
$s_+$ and $s_-$ respectively, the solutions for these contours
are valid for $x>0$.  The solution corresponding to the contour
$C_4$ is valid for $x<0$.
The unmarked regions are directions in which the contour must
asymptote for the integral to converge.  The $\times$'s are
singularities of the integrand and the wavy line is a branch cut.}
\label{contour5}
\end{figure}

Because the details of computing the other n-mode solution discussed
above are essentially the
same as discussed in subsections \ref{subapprox} and \ref{subwkb}, we 
shall only sketch the computation.  We 
first solve the mode equation (\ref{ODE}) (with the plus sign) about 
the horizon
by the method of Laplace transforms exactly as before.  The only
difference is that the contour of integration must be changed
so as to satisfy the boundary conditions that the solution reduce
to a plane wave with wavevector $k_-$ inside the horizon where
$v(x)$ is essentially constant.  A straightforward computation shows
that the contour $C_4$ shown in figure \ref{contour5} does 
the job, i.e., evaluating the
contour integral (\ref{contint}) over $C_4$ and propagating
the solution deeper inside the black hole by the WKB approximation
to the constant $v(x)$ region 
shows that the solution is
\begin{equation}
\phi_4(x) \approx 2 e^{\pi \omega/\kappa} \sinh(\pi \omega/\kappa)
\Gamma(-i \omega/\kappa) \phi_{-}(x),
\end{equation}
which is the boundary condition we want.

To evaluate the solution outside the horizon, we deform the
contour $C_4$ into $C_2$ and $C_5$ as shown in figure (\ref{contour5}).
These we can evaluate by the method of steepest descents.  In
fact $C_2$ is exactly the same as the contour $C_2$ before, see
figure \ref{contour2}.
$C_5$ is the same as $C_1$ before (see figure \ref{contour2}), except 
that it lies
on a different Riemann sheet, so only the overall scale
changes.  The complete solution outside the horizon after
being propagated out to the constant $v(x)$ region by the WKB
approximation is
\begin{equation}
\phi_2(x)+\phi_5(x) \approx e^{-i \pi/4} 
e^{\pi \omega /(2 \kappa)} \, \sqrt{2 \pi \kappa} \,
(\phi_{+}(x) -i e^{\pi \omega/\kappa}
\phi_{-}(x)).
\end{equation}
Combining this with $\phi_4(x)$ produces the connection formula
\begin{equation}
2 e^{\pi \omega/\kappa} \sinh(\pi \omega/\kappa)
\Gamma(-i \omega/\kappa) \phi_{-}(x)
\leftrightarrow
e^{-i \pi/4} \, 
e^{\pi \omega /(2 \kappa)} \, \sqrt{2 \pi \kappa} \,
(\phi_{+}(x) -i e^{\pi \omega/\kappa}
\phi_{-}(x))
\label{nmode2}
\end{equation}
where the left-hand-side refers to $x<0$ and the right-hand-side to
$x>0$.
This is the second of the n-mode solutions, the first is
given in (\ref{wkbform}).

To obtain the s-mode solutions, we add the n-mode solutions given
in (\ref{wkbform}) and (\ref{nmode2}) with the correct relative coefficient so
that either the $\phi_{-}(x)$ mode or the $\phi_{+}(x)$ mode
cancels at large positive $x$, this produces the connection
formulae
\begin{eqnarray}
e^{\pi \omega/\kappa} \phi_{-}(x) & \leftrightarrow &
\phi_{+s}(x) + e^{-i3 \pi/4} \frac{\sqrt{2 \pi \kappa}}{\Gamma(-i \omega
/\kappa)} \phi_{-}(x) \\ \label{negcon}
e^{-\pi \omega/\kappa} \phi_{-}(x) & \leftrightarrow &
\phi_{+s}(x) -
e^{-i \pi/4} e^{-3\pi \omega/(2 \kappa)}
\frac{\sqrt{2 \pi \kappa}}{\Gamma(-i \omega
/\kappa)} \phi_{+}(x)
\label{poscon}
\end{eqnarray}
where again the left-hand-side refers to negative $x$ and the
right-hand-side to positive $x$.
Noting that the annihilation operator associated with the modes
$\phi_{+}(x)$ and $\phi_{-}^{*}(x)$ at large
positive $x$ (in the time dependent picture these would
be the early time ingoing, positive free fall frequency wavepackets)
both annihilate the
free fall vacuum $| {\rm ff} \rangle$, we derive the following relations
as in (\ref{stateeqn}),
\begin{eqnarray}
(a^{\dag}(\phi_{+s}) - e^{\pi \omega/\kappa} a(\phi_{-}^*))| 
{\rm ff} \rangle =0 \\
(a^{\dag}(\phi_{-}^*) - e^{\pi \omega/\kappa} a(\phi_{+s}))| 
{\rm ff} \rangle =0. 
\label{anniheqns}
\end{eqnarray}
These two relations completely determine the $\phi_{+s}$ and
$\phi_{-}^{*}$ content of the free fall vacuum
for the Killing frequency $\omega$.  Using these relations it is
simple to show that the free fall vacuum is a thermal state at the
Hawking temperature, exactly as with the ordinary wave equation, 
\cite{Wald75}.
 
\section{Approximate solutions to the superluminal equation}
\label{sup}

Since the calculations involved in solving the mode equation
(\ref{ODE}) with the superluminal operator $\m$ are
virtually identical to those given above for the subluminal
operator $\p$, we shall be brief.  
Computing the approximate solutions outside the horizon by the
WKB approximation proceeds exactly as before.
The main difference compared to the subluminal case is that
we $are$ now interested in WKB solutions both
inside and outside the horizon.  
Outside the horizon,
the relevant\footnote{There are other linearly independent solutions
both inside and outside the horizon, but they will not be needed
in this calculation.} WKB mode is
\begin{equation}
\phi_{+s}(x) \approx e^{i \omega \int dx/(1 + v(x))},
\label{WKBpsm}
\end{equation}
and inside the relevant WKB modes are
\begin{equation}
\phi_{\pm}(x) \approx (-1+v^{2}(x))^{-3/4} e^{\pm i \int dx \, \sqrt{
-1 + v^2(x)}} e^{i \omega \int dx \, v(x)/(1 - v^2(x))}.
\label{WKBpmm}
\end{equation}
As before, it is straightforward to show that these approximate solutions
break down around the horizon, but far enough outside the horizon
they are valid.
To find the appropriate connection formula for these solutions, we now
find an approximate solution across the horizon.

Solutions about the horizon
can be obtained by again linearizing $v(x)$ as in (\ref{vexpand})
and solving the
resulting approximate mode equation by the method of Laplace
transforms.
The linearized equation is just (\ref{ODE1}) with a minus sign inserted
before the fourth derivative term.  Writing the solution as a Laplace
transform as in (\ref{laplace}) and substituting into the 
equation produces the $s-$space ODE (\ref{sODE})
with $s^4 \rightarrow - s^4$.  This equation is again trivial to solve.
One finds upon writing the solution in the form
\begin{equation}
\phi(x) = \int_{C} ds \, g(s) e^{x f(s)}
\label{contint2}
\end{equation}
that
\begin{equation}
f(s) = s - \frac{1}{2 \kappa x} \left( \frac{s^3}{3} - 
\frac{i \omega (i \omega - \kappa)}{s} \right)
\label{fofsm}
\end{equation}
and
\begin{equation}
g(s) = s^{-1 - i \omega/\kappa}.
\end{equation}

\begin{figure}[hbt]
\centerline{
\psfig{figure=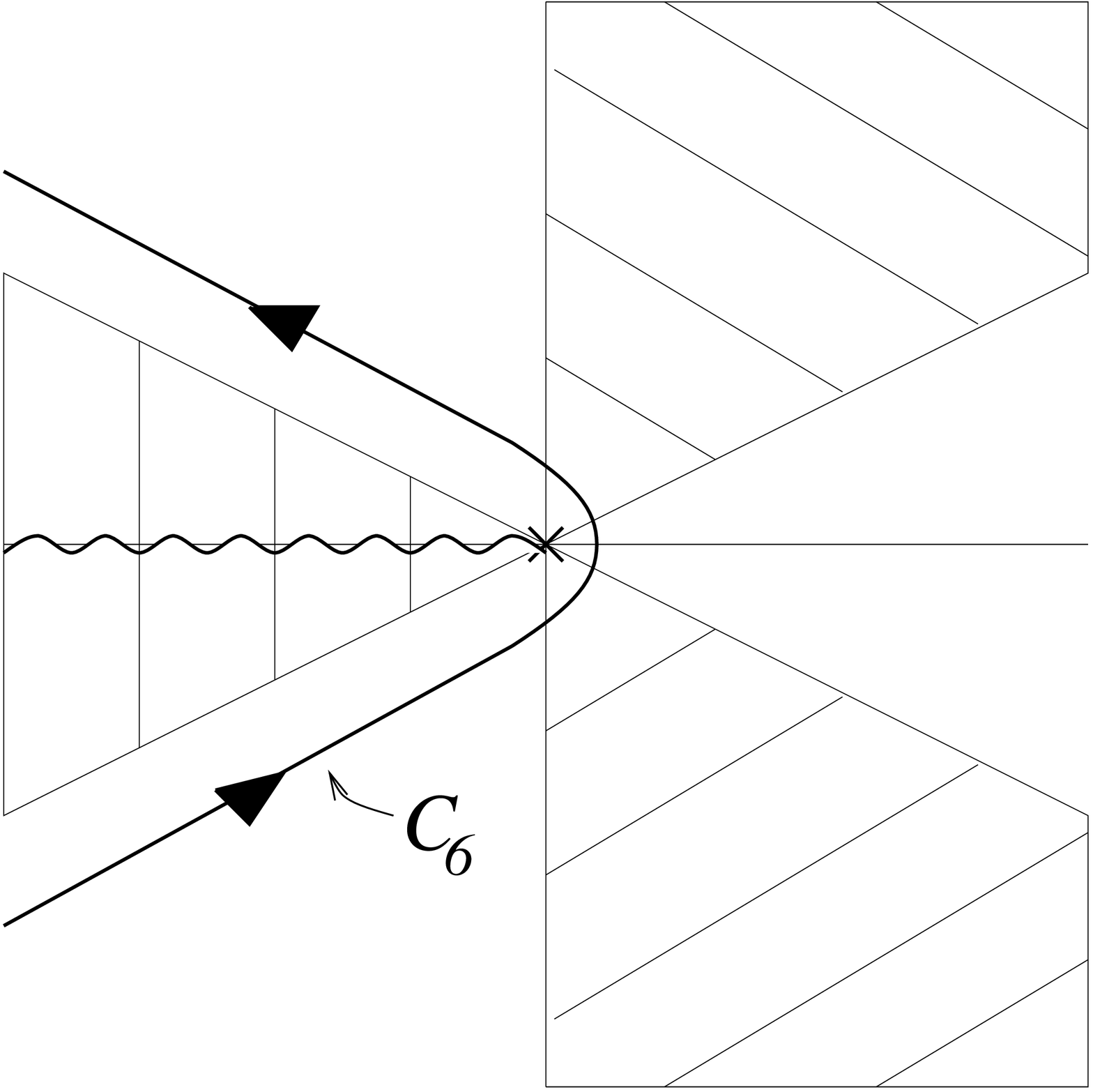,angle=-90,height=8cm}}
\caption{\small Diagram of the steepest descent contour $C_{6}$.
The unmarked regions are directions in which the contour must
asymptote for the integral to converge.  The $\times$'s are
singularities of the integrand and the wavy line is a branch cut.}
\label{contour3}
\end{figure}

Before evaluating (\ref{contint2}), first note that
at large $|s|$, the integral is dominated by $\exp(-s^3/(6 \kappa))$, and
therefore for the integral to converge, the contour must asymptote to
a region where $Re(s^3) > 0$.  Writing $s=r e^{i \theta}$, these regions
are given by
\begin{eqnarray}
Region \, 1 & \leftrightarrow & \frac{-\pi}{6} < \theta < \frac{\pi}{6} \\
Region \, 2 & \leftrightarrow & \frac{\pi}{2} < \theta < \frac{5 \pi}{6} \\ 
Region \, 3 & \leftrightarrow & \frac{7 \pi}{6} < \theta < \frac{3 \pi}{2},
\label{regions2}
\end{eqnarray}
and are the unmarked regions in figure \ref{contour3}.

To evaluate the contour integral (\ref{contint2}), recall from section
\ref{computing} that
our boundary conditions are specified outside the horizon and
state that the solution must reduce to a plane wave with wavevector
$k_{+s}$ in the constant
$v(x)$ region.
This solution does not correspond to a saddle point
because those solutions are either exponentially
growing or decaying.  It is not hard to guess what contour we need though,
given our past experience with the subluminal dispersion relation.  If 
we take a contour $C_6$
that encircles the branch cut and asymptotes to regions 2 and 3 
(\ref{regions2}), see
figure \ref{contour3}, we get a contour very similar to the contour $C_3$
in figure \ref{contour2}.  The approximations that went into evaluating
that contour also work here.  The result is exactly $-\phi_3(x)$ of
(\ref{phi3}).  This solution is just $\phi_{+s}(x)$ (\ref{WKBpsm}) up
to a multiplicative constant, and therefore the contour $C_{6}$
produces the correct boundary condition at $x \gg 0$.

\begin{figure}[hbt]
\centerline{
\psfig{figure=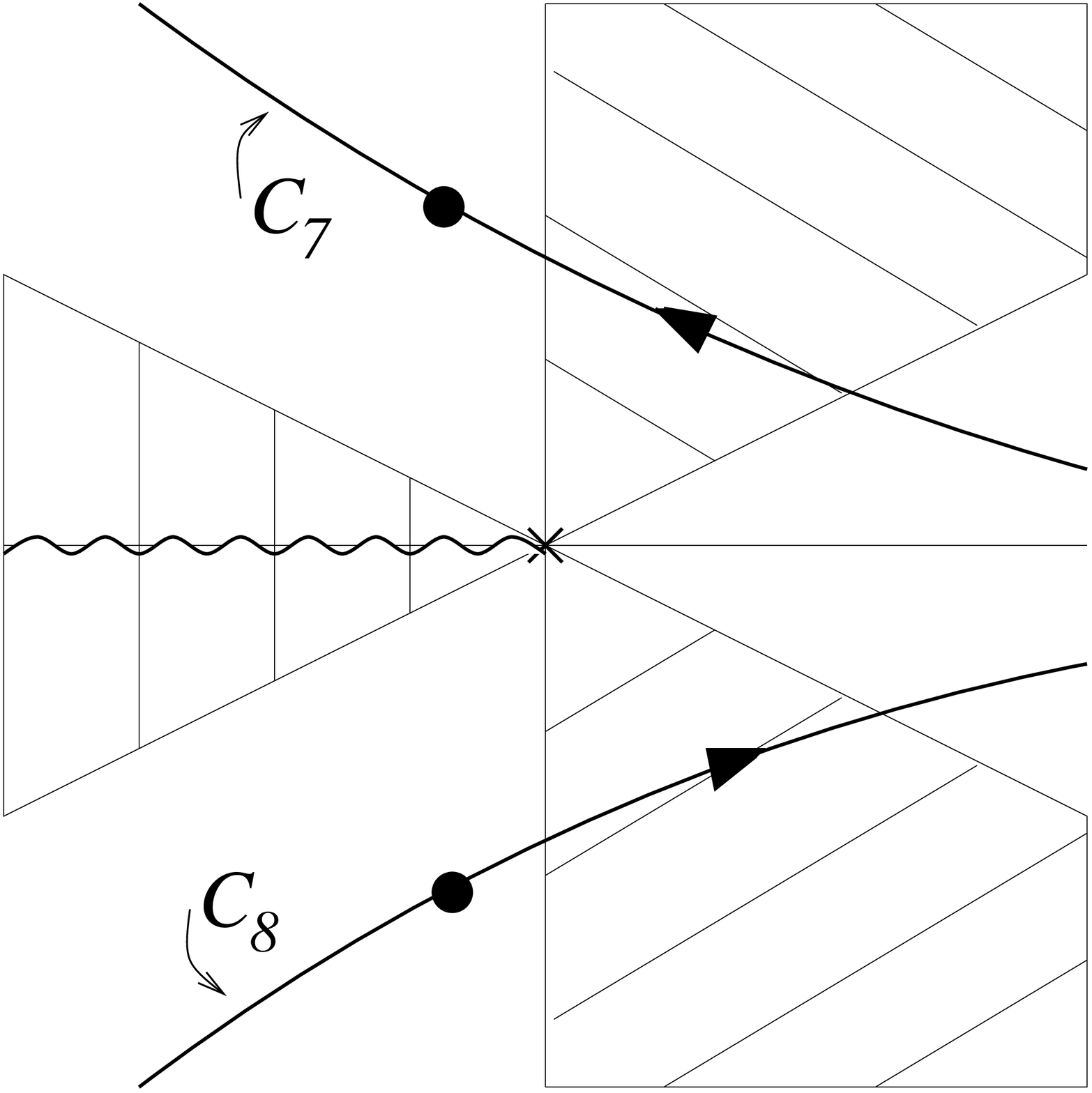,angle=-90,height=8cm}}
\caption{\small Diagram of the steepest descent contours, $C_{7}$
and $C_{8}$, through the saddle points $s_+$ and $s_-$ respectively.
The unmarked regions are directions in which the contour must
asymptote for the integral to converge.  The $\times$'s are
singularities of the integrand and the wavy line is a branch cut.}
\label{contour4}
\end{figure}

To evaluate (\ref{contint2}) for $x \ll 0 $ we use the steepest
descents approximation.  The saddle points in the integrand of
(\ref{contint2}) are given by
\begin{equation}
s_{\pm} \approx \pm i \sqrt{2 \kappa |x|},
\end{equation}
and the steepest descent contours must pass through these points in the
directions $-\pi/4$ to $3 \pi/4$ for $s_{+}$ and $\pi/4$ to $5 \pi/4$
for $s_{-}$.  The contours, $C_{7}$ and $C_{8}$, therefore
must asymptote to regions 1 and 2 and regions 2 and 3 respectively, see
figure \ref{contour4}.  Furthermore, $C_{7} + C_{8}$ is deformable to
$C_{6}$, and therefore is the contour we want.  Evaluating (\ref{contint2})
over 
$C_{7} + C_{8}$ by the steepest descents approximation and expressing
the result in terms of the WKB solutions (\ref{WKBpmm}) results in
\begin{equation}
\phi_{7} +\phi_{8}(x)  \approx  e^{-i \pi/4} \sqrt{2 \pi \kappa}
(e^{- \pi \omega/{2 \kappa}}
\phi_{-}(x) 
-i e^{ \pi \omega/{2 \kappa}}
\phi_{+}(x)).
\end{equation} 
We now have the complete solution for all $x$, which can be displayed
as the connection formula
\begin{equation}
e^{-i \pi/4} \sqrt{2 \pi \kappa}(
e^{- \pi \omega/{2 \kappa}}
\phi_{-}(x) 
-i e^{ \pi \omega/{2 \kappa}}
\phi_{+}(x)) \leftrightarrow  \frac{-i 2 \pi}{\Gamma(1+i\omega/\kappa)}
\phi_{+s}(x)
\label{supcon1}
\end{equation}
where the left-hand-side refers to $x<0$ and the right-hand-side to $x>0$.
Evaluating (\ref{supcon1}) at $x \gg 0$
and $x \ll 0$ allows us to pull off the coefficients $c_{+s}(\omega)$
and $c_{-}(\omega)$.  Substituting into (\ref{No}) again produces a thermal
spectrum at the Hawking temperature.

As before we have made a number of approximations to arrive at this
result.  Collecting these approximations together, we can compute the
range of validity of these results.  However, because the difference
between the solutions for the subluminal equation and the superluminal
equation is only the change of a few signs in the end, then the 
difference between the WKB and Laplace transform solutions for
the superluminal case is essentially the same as the difference
between the WKB and Laplace transform solutions for the subluminal
case.  Therefore the constraints on the range of validity of
the solutions in the
parameters $\kappa$, $x$, and $\omega$ are the same as in the
subluminal case.

\subsection{Computing the quantum state}
\label{supstate}

For the superluminal equation we can as well compute the decomposition
of the free fall vacuum in terms of particle states as seen by
late time observers.  We define the out Hilbert space as before, i.e.,
we take it to be a tensor product of Hilbert spaces inside and outside
the horizon respectively.  Outside the horizon we define the Hilbert space
using Killing frequency, and inside we use free fall frequency (as before
we take a $v(x)$ that asymptotes to a constant smaller than $-1$ behind
the horizon).

To compute the decomposition of the free fall vacuum, we again look
for mode solutions which when summed together produce an early
time positive free fall frequency packet propagating toward the
horizon (but now located behind the horizon).  Around the horizon
this packet scatters (mode converts) into a pair of packets,
a reflected packet which propagates deeper inside the black hole
to the constant $v(x)$ region and a transmitted packet which 
propagates across the horizon out to the constant $v(x)$ region.
In this picture the transmitted packet corresponds to the Hawking
particle and the reflected packet to the partner.  From such
a solution we could obtain an equation for the free fall vacuum
analogous to (\ref{stateeqn}).

To compute the s-modes (those needed to compute the state) we again
first compute the n-modes (those needed to compute number expectation
values).   We have already computed one n-mode given by
(\ref{supcon1}).  Recall that in the time dependent picture this
corresponds to a pair of positive and negative free fall frequency
packets propagating toward to the horizon (and located inside
the black hole) with just the right relative coefficient that
the entire packet propagates across the horizon out to the
constant $v(x)$ region.
The other n-mode therefore correponds in the time dependent picture
to a pair of positive and negative free fall frequency packets
propagating toward the horizon with the right relative coefficient such
that the entire packet is reflected and propagates deep inside the
black hole to where $v(x)$ is constant.  The appropriate
mode solution therefore must decay $outside$ the horizon.

Computing this mode solution involves the same techniques used already
many times, so we shall simply quote the result.  The connection
formula expressed in terms of the WKB solutions is
\begin{eqnarray}
(e^{-i \pi/4} \sqrt{2 \pi \kappa} (-i e^{\pi \omega/(2 \kappa)} \phi_{+}(x)
+ e^{3 \pi \omega/(2 \kappa)} \phi_{-}(x)) \nonumber \\ 
+i \,e^{\pi \omega/\kappa} \frac{2 \pi}{\Gamma(1+i\omega/\kappa)} \phi_{-m}(x))
\leftrightarrow e^{-i3 \pi/4} e^{\pi \omega/\kappa} \phi_{+}(x)
\label{supcon2}
\end{eqnarray}
where the left-hand-side refers to $x<0$ and the right-hand-side
to $x>0$.  The solution $\phi_+(x)$ for positive $x$ decays exponentially
with increasing $x$, and therefore satisfies the boundary conditions.

To compute the s-modes we take linear combinations of the n-modes
(\ref{supcon1}) and (\ref{supcon2}) to eliminate either
$\phi_+(x)$ or $\phi_-(x)$ behind the horizon.  This results in 
the connection formulae
\begin{eqnarray}
\phi_-(x) & + & i N e^{\pi \omega/\kappa} \phi_{-m}(x) \leftrightarrow
i N \phi_{+s}(x) \\
\phi_+(x) & + &  N \phi_{-m}(x) \leftrightarrow
N  e^{\pi \omega /\kappa} \phi_{+s}(x)
\end{eqnarray}
where 
\begin{equation}
N = e^{i \pi/4} e^{- \pi \omega/(2 \kappa)} \frac{\pi}{\sqrt{2 \pi \kappa} \,
\sinh(\pi \omega/\kappa) \, \Gamma(1+i\omega/\kappa)}.
\end{equation}
These relations are enough to carry out the decomposition of the 
free fall vacuum as discussed in subsection \ref{substate}.  
In particular the method of obtaining the equations (\ref{anniheqns})
on the free
fall vacuum for the subluminal dispersion relation follows
exactly in this case as well, with the replacement
of $\phi_-$ in the subluminal case by $\phi_{-m}$ in
the superluminal case.

\section{Conclusions}
\label{conclusions}

We have considered two different modifications of the wave equation in a
black hole spacetime, one producing subluminal propagation of high 
frequency modes and the other superluminal propagation of high frequency
modes.  We have shown that both equations give rise to exactly a
thermal spectrum of radiation from a black hole to leading order
in an expansion in powers of $1/k_0$.  It is natural to try
to push the analysis further to obtain a correction term to the outgoing
flux.  We immediately run into the following difficulty though.  In
obtaining an approximate solution to (\ref{ODE}) about the horizon,
we actually solved instead just the linearized equations (\ref{ODE1}).
To obtain a better approximate solution about the horizon, we need
a better approximation to (\ref{ODE}).  We could, for instance, keep
higher order terms in $x$ when expanding $v(x)$ and $v^{\prime}(x)$.  
If we try to solve the resulting equation by the Laplace transform
method, we find that we get a higher order ordinary differential equation in
the Laplace transform variable $s$.  In other words, the $s$-space
equation is really not much better than the original $x$-space
equation.

An important assumption made in deriving the thermal radiation for the
superluminal equation of motion
was that positive free fall frequency modes, located behind
the horizon, were in their ground state.  Clearly we don't know a priori
whether this is the physically correct quantum state condition.  In 
principle we would have to begin with a quantum state which evolves
into a black hole, and then ask if these modes actually are in their
ground state.  This requires quantum gravity.  A more realistic
problem to tackle at this time is simply to ask where these modes
came from in a semiclassical approximation.  One would guess naively
from the singularity.  Recent investigations, \cite{Jac, MarHor}, have
shown; however, that for certain charged black holes it is possible that
these modes simply reflect outside the singularity and become
ingoing modes, backward in time.  This would have important implications
because it would mean that the Hawking radiation, even for an eternal
black hole, would originate from ingoing modes, and therefore we
would not have the infinite degrees of freedom problem, \cite{Jac}.

We end by noting that the subluminal model (and possibly the 
superluminal model as well) considered in this paper suffers
from the ``stationarity puzzle''.  If we try to propagate the
outgoing modes backward in time all the way out to infinity, where
$v(x)$ goes to zero, then there can be no particle creation by
conservation of Killing frequency.  One way out of this problem
is to introduce time dependence into the equation of motion (perhaps
via backreaction) to destroy the Killing symmetry.  A step in this
direction is to put the ordinary wave equation on a spatial
lattice, this has the advantage of introducing naturally a
short distance cutoff and at the same time destroying the
Killing symmetry (for discretizations of most spatial
coordinates).  Such a model is currently being investigated \cite{CJ2}
by techniques similar to those described in this paper.

\section{Acknowledgements}

I would like to thank Ted Jacobson for many valuable conversations
and suggestions on a draft of this paper, and also Jorma Louko
for helpful discussions.  This work was supported in part by NSF grant
PHY94-13253.

\end{document}